\documentclass[11pt]{article}

\usepackage{amsmath, amsfonts, mathrsfs,latexsym,color,array}

\usepackage{cite}
\usepackage{url}
\usepackage{ntheorem}

\newcommand{\Nnmo}{N^{n-1}}
\newcommand{\Snmo}{\Sp^{n-1}}

\definecolor{bluem}{rgb}{0,0,0.5}

\definecolor{mycolor}{cmyk}{0.5,0.1,0.5,0}
\definecolor{michel}{rgb}{0.5,0.9,0.9}

\definecolor{turquoise}{rgb}{0.25,0.8,0.7}
\definecolor{bluem}{rgb}{0,0,0.5}

\definecolor{MDB}{rgb}{0,0.08,0.45}
\definecolor{MyDarkBlue}{rgb}{0,0.08,0.45}

\definecolor{MLM}{cmyk}{0.1,0.8,0,0.1}
\definecolor{MyLightMagenta}{cmyk}{0.1,0.8,0,0.1}

\definecolor{HP}{rgb}{1,0.09,0.58}

\newcommand{\ourU}{\mathbb U}

\newcommand{\Ric}{\mathrm{Ric}}

\newcommand{\zGamma}{\mathring \Gamma}

{\catcode `\@=11 \global\let\AddToReset=\@addtoreset}
\AddToReset{equation}{section}

\newcommand{\sqrtV}{\zN}

\newcommand{\zlambda}{\mathring \lambda}
\newcommand{\zm}{\mathring m}

\newcommand{\zN}{\mathring N}
\newcommand{\zM}{\mathring M}
\newcommand{\zG}{{\mathring \Gamma}}
\newcommand{\mcD}{{\mycal D}}
\newcommand{\zD}{{\,\,\,\mathring{\!\!\! \mcD}}{}}

\newtheorem{Theorem} {\sc  Theorem\rm} [section]

\newtheorem{Lemma} [Theorem] {\sc  Lemma\rm}
\newtheorem{Proposition} [Theorem] {\sc  Proposition\rm}

\newtheorem{theorem}[Theorem]{\sc  Theorem\rm}

\theorembodyfont{\upshape}

\newtheorem{remark}[Theorem]{\sc Remark\rm}

\newcommand{\beqar}{\begin{deqarr}}
\newcommand{\eeqar}{\end{deqarr}}

\newcommand{\beaa}{\begin{eqnarray*}}
\newcommand{\eeaa}{\end{eqnarray*}}

\newcommand{\bel}[1]{\begin{equation}\label{#1}}
\newcommand{\bea}{\begin{eqnarray}}
\newcommand{\bean}{\begin{eqnarray}\nonumber}
\newcommand{\beal}[1]{\begin{eqnarray}\label{#1}}
\newcommand{\eea}{\end{eqnarray}}
\newcommand{\eeal}[1]{\label{#1}\end{eqnarray}}

\newcommand{\Eq}[1]{Equation~\eq{#1}}

\newcommand{\qed}{\hfill $\Box$ \medskip}
\newcommand{\proof}{\noindent {\sc Proof:\ }}
\newcommand{\be}{\begin{equation}}
\newcommand{\eeq}{\end{equation}}
\newcommand{\ee}{\end{equation}}
\newcommand{\beqa}{\begin{eqnarray}}
\newcommand{\eeqa}{\end{eqnarray}}
\newcommand{\beqan}{\begin{eqnarray*}}
\newcommand{\eeqan}{\end{eqnarray*}}
\newcommand{\ba}{\begin{array}}
\newcommand{\ea}{\end{array}}

\newcommand{\hyp}{\mycal S}

\DeclareFontFamily{OT1}{rsfs}{} \DeclareFontShape{OT1}{rsfs}{m}{n}{
<-7> rsfs5 <7-10> rsfs7 <10-> rsfs10}{}
\DeclareMathAlphabet{\mycal}{OT1}{rsfs}{m}{n}

\newcommand{\ptc}[1]{\mnote{{\bf ptc:}#1}}

\newcommand{\Sp}{\mathbb S}
\newcommand{\R}{\mathbb R}

\newcommand{\bit}{\begin{itemize}}
\newcommand{\eit}{\end{itemize}}

\newcommand{\g}{\gamma}
\newcommand{\tg}{{\tilde g}}
\newcommand{\ve}{{\varepsilon}}

\newcommand{\trg}{{\mbox{\rm tr}_g}}

\newcounter{shownewstuffflag}

\newcommand{\startnewstuff}{\ifnum\value{shownewstuffflag}>0\color{blue}\fi}
\newcommand{\finishnewstuff}{\ifnum\value{shownewstuffflag}>0\color{black}\fi}

\newcounter{oldeq}

\newcounter{mnotecount}[section]

\newcommand{\mnote}[1]{}

\newcommand{\rmnote}[1]{}

\def\beq{\begin{equation}}
\def\eeq{\end{equation}}

\newcommand{\zh}{\,\,{\mathring{\!\!h}}}
\newcommand{\zg}{{\mathring g}}

\newcommand{\eq}[1]{(\ref{#1})}

\begin{document}
\title{Singular Yamabe metrics and initial data 
with \emph{exactly} Kottler--Schwarzschild--de Sitter ends}
\author{Piotr T. Chru\'sciel
\\  LMPT, F\'ed\'eration Denis Poisson, Tours \\ Mathematical Institute and Hertford College,
Oxford
\\
\\
Daniel Pollack
\\ University of Washington}


\maketitle

\abstract{ We construct  large families of initial data sets for the
vacuum Einstein equations with positive cosmological constant which
contain \emph{exactly Delaunay ends}; these are non-trivial initial
data sets which coincide with those for the
Kottler--Schwarzschild--de Sitter metrics in  regions of infinite
extent. {}From the purely Riemannian geometric point of view, this
produces complete, constant positive scalar curvature metrics with
exact Delaunay ends which are not globally Delaunay. The ends can be
used to construct new compact initial data sets via gluing
constructions.  The construction provided applies to more general
situations where the asymptotic geometry may have
\emph{non-spherical} cross-sections consisting of Einstein metrics
with positive scalar curvature.}

\section{Introduction} \label{Sintro}

There exists very strong evidence suggesting that we live in a world
with strictly positive cosmological constant
$\Lambda$~\cite{WoodVasey:2007jb,Riess:2006fw}. This leads to a need
for a better understanding of the space of solutions of Einstein
equations with $\Lambda>0$. The most general method available for
constructing such solutions proceeds by solving a Cauchy
problem~\cite{BartnikIsenberg,FriedrichRendall,CBY}. In view of the
general relativistic constraint equations this, subsequently,
requires understanding the corresponding collection of initial data
sets. In particular one is led to the question of boundary
conditions satisfied by the fields. When $\Lambda$ vanishes a
natural set of boundary conditions arises from the obvious model
solution -- the Minkowski space-time.  A tempting further
restriction is then the requirement of a well defined and finite
total mass,   leading to a well understood set of asymptotic
boundary conditions~\cite{Bartnik86,Chremark,Omurchadha86,BOM}. When
$\Lambda>0$ the question of asymptotic conditions seems to be much
less clear cut. One wants to consider a class of space-times which
includes all solutions of physical interest. Until there is
overwhelming evidence to the contrary, ``physical interest" should
carry a notion of ``non-singular". The simplest possibility, widely
adopted, is to assume that the Cauchy surface $\hyp$ is a compact
manifold without boundary. However, an appealing more general way of
ensuring regularity of the initial data is to suppose that
$(\hyp,g)$ is a complete Riemannian manifold. One would then like to
understand the space of solutions of those general relativistic
constraint equations with $(\hyp,g)$ -- complete.

An interesting class of asymptotic models for such initial data has
already been explored in the mathematical literature, the time
symmetric initial data provided by the
\emph{Delaunay}\footnote{These are also often called \emph{Fowler}
solutions; see \S \ref{sDDm} for further remarks on the  history and
choice of terminology used here.}
metrics~\cite{SchoenCatini,Fowler1,Fowler2}. These describe the
family of complete rotationally symmetric, conformally flat
metrics  with constant positive scalar curvature, and are in fact
well known to general relativists as the time-symmetric slices of
the Kottler--Schwarzschild--de Sitter
solutions~\cite{GibbonsHawkingCEH,Kottler} (however the connection
between these two subjects has apparently not been previously
noted.  In the Riemannian geometric context, the Delaunay metrics
form the local asymptotic model for isolated singularities of
locally conformally flat constant positive scalar curvature
metrics~\cite{CGS,P93,ChenLin95,KMPS} (in dimensions $n\leq5$ this
also holds in the non-conformally flat setting\cite{Marques}). The
known results concerning the existence of complete constant positive
scalar curvature metrics with \emph{asymptotically Delaunay}
ends~\cite{S88,P93,MPU1,MPU2,MPa,MPo,KMPS,Rat,Byde} may thus be
reinterpreted, via their space-time development, as the existence of
space-times satisfying the Einstein field equations with a positive
cosmological constant which have asymptotically
Kottler--Schwarzschild--de Sitter ends.

The object of this work is to point out that every
constant positive scalar curvature (CPSC) asymptotically Delaunay
metric is naturally accompanied by a CPSC metric with an
\emph{exactly Delaunay} end, and moreover these metrics may be
choosen to coincide away from the end in question.
Such metrics  are of interest in general relativity for at least
four reasons:

\begin{enumerate}
\item They provide, via their maximal development, a large class of space-times
satisfying the Einstein field equations
with a positive cosmological constant with exactly controlled
geometry in the asymptotic regions; in fact the space-time
development is explicitly known in the domain of dependence of the
Delaunay regions.

\item They demonstrate that the special horizon behavior, with alternating cosmological and event
horizons, which is exhibited by the Kottler--Schwarzschild--de
Sitter space-time,   occurs in large classes of non-stationary
solutions.

\item Any two metrics which carry exactly Delaunay ends with identical mass (Delaunay) parameters
may be glued together using obvious identifications on the
ends. (A more difficult \emph{end-to-end} asymptotic gluing theorem of this sort was
established by Ratzkin~\cite{Rat}, however with exactly Delaunay
ends this construction is effortless.) Thus  Delaunay ends can
easily be used as bridges to create wormholes, or to make connected
sums of initial data sets. Wormhole constructions are already known
to be possible by completely different techniques~\cite{CIP} in the
setting of a non-positive cosmological constant. Here we provide
such a construction for  positive cosmological constants,  with the
added bonus of an explicit knowledge of the space-time development
in the domain of dependence of the middle part of the connecting
neck, which may be of arbitrary (quantized by multiples of the period of the
exact Delaunay metric) length.

 \item The asymptotically Delaunay metrics are uniquely characterised by a
simple geometric criterion~\cite{KMPS,Marques}, see
Section~\ref{sDDm} below.

\end{enumerate}

A natural setting for our considerations is provided by the
\emph{generalised Kottler metrics} and \emph{generalised Delaunay
metrics}, as described in Sections~\ref{ssgKm} and \ref{ssgDm}
below. Our gluing construction applies in this more general setting.

In an accompanying paper~\cite{ChDelayAH}, by one of us (PTC) and
Erwann Delay, analogous constructions are carried out with a
negative cosmological constant. With hindsight, within the family of
Kottler metrics with $\Lambda \in \R$, the gluing in the current
setting is the easiest, while that in~\cite{ChDelayAH} is the most
difficult. This is due to the fact that for $\Lambda>0$, as
considered here, one deals with one linearised operator with a
one-dimensional kernel; in the case $\Lambda=0$ the kernel is
$(n+1)$--dimensional; while for $\Lambda<0$ one needs to deal with a
one-parameter family of operators with $(n+1)$--dimensional kernels.

\medskip
\noindent {\sc Acknowledgements:} DP  would like to thank Mihalis
Dafermos for first raising the question of whether space-times with
Kottler--Schwarzschild--de Sitter horizon behavior exist more generally, and
Frank Pacard for a number of illuminating discussions.

\section{Kottler--Schwarzschild--de Sitter space and metrics of constant positive scalar curvature
with asymptotically Delaunay ends}
\label{se:intro}

In this section we review some results concerning the
Kottler--Schwarzschild--de Sitter space and CPSC metrics which are
asymptotically Delaunay.  In order to fix notations and conventions
we start with some standard facts.

Recall that initial data for the Einstein field equations with a
cosmological constant $\Lambda$ on an $n$-dimensional manifold $M$
consist of a pair $(g, K)$ consisting of a Riemannian metric $g$ on
$M$ and a symmetric 2-tensor $K$ satisfying the vacuum constraint
equations
\begin{eqnarray}
\label{ce1}
R(g) -(2\Lambda+ |K|_g^2-(\trg  K)^2)= 0\\
D_i(K^{ij}-\trg  K g^{ij})=0 \label{ce2}
\end{eqnarray}
where $R(g)$ is the scalar curvature (Ricci scalar) of the  metric $g$.  If one considers time-symmetric initial data, for which
$K\equiv 0$, then these equations reduce to the requirement that $g$ has constant scalar curvature $R(g)=2\Lambda$.
Here we restrict to the case where $\Lambda$ is positive, and note that the  normalization $\Lambda=\frac{n(n-1)}{2}$
corresponds to $R(g)=n(n-1)$, the scalar curvature of the standard sphere of radius one in
${\mathbb R}^{n+1}$.

\subsection{Kottler--Schwarzschild--de Sitter metrics}
 \label{KSdS}

The Kottler--Schwarzschild--de Sitter space-time~\cite{Kottler}
metric in $n+1$ dimensions, with cosmological constant $\Lambda>0$
and mass $m\in \R$ may be written as
 \ptc{the circle over the $h$ in $\zh$ is not placed correctly here and elsewhere (to change throughout the paper)}
\bel{Schwm}
 ds^2 = -V dt^2 + V^{-1} dr^2 + r^2\zh ,\quad
\mbox{where}\quad V = V(r) = 1-\frac{2m}{r^{n-2}} -\frac{
r^2}{\ell^2} \;,
 \ee
where $\ell>0$ is   related to the cosmological constant $\Lambda$
by the formula $2\Lambda=n(n-1)/\ell^2$, while $\zh$ denotes the
standard metric on the unit $(n-1)$-sphere in $\R^n$. To avoid a
singularity lying at finite distance on the level sets of $t$ we
will assume $m>0$. Equation \eq{Schwm} provides then a spacetime
metric satisfying the Einstein equations with cosmological constant
$\Lambda>0$ and with well behaved spacelike hypersurfaces when one
restricts the coordinate $r$ to an interval $(r_b, r_c)$ on which
$V(r)$ is positive; such an interval exists if and only if
\bel{extreme} \left(\frac{2}{(n-1)(n-2)}\right)^{n-2}\Lambda^{n-2}
m^2 n^2 <1
 \;.
\ee
When $n=3$ this corresponds to the condition that
$9m^2\Lambda<1$, and the case of equality is referred to as the
extreme Kottler--Schwarzschild--de Sitter  space-time (for which the
coordinate expression (\ref{Schwm}) is no longer valid).  In the
limit where $\Lambda$ tends to zero with $m$ held constant, the
space-time metric approaches the Schwarzschild metric with mass $m$,
and in the limit where $m$ goes to zero with $\Lambda$ held constant
the metric tends to that of the de Sitter space-time with
cosmological constant $\Lambda$.

The breakdown of the coordinate description above at the horizons
$r=r_b$ and $r=r_c$ can be handled by taking
extensions~\cite{GibbonsHawkingCEH,BH}: In fact, the
Kottler--Schwarzschild--de Sitter metric admits an analytic
extension (analogous to the Kruskal extension of the Schwarzschild
metric) as an $r$--periodic metric on $(t, r, \theta)\in
\R\times\R\times \Sp^{n-1}$.  This is most easily seen via the
associated conformal Carter-Penrose
diagrams~\cite{GibbonsHawkingCEH}. The time-symmetric slice $t=0$ of
the (extended) Kottler--Schwarzschild--de Sitter metrics are thus a
one-parameter family (parameterized by their mass $m$) of periodic,
spherically symmetric, metrics on $\R\times \Sp^{n-1}$  with
constant positive scalar curvature $R=2\Lambda$.

Finally note that, due to the spherical symmetry, each of these
metrics is conformally flat.

 \subsubsection{Extreme limit}
 \label{sssXl}
It is of some interest to enquire what happens when $m\to\mathring
m$, where $\mathring m$ denotes the values at which equality is
achieved in \eq{extreme}. In this limit $r_b$ and $r_c$ coalesce to
a single value which we will denote by $\mathring r$. From the
space-time point of view the situation is the following: recall that
the Carter-Penrose diagram for the maximally extended KSdS
space-times with $0<m<\mathring m$ is built out of diamond shaped
regions corresponding to $r_b<r<r_c$, where the Killing vector
$\partial_t$ is time-like, and of triangle shaped (either upright,
or upside-down) regions where $\partial_t$ is spacelike{~\cite{GibbonsHawkingCEH}}.
After
passing to  the limit $m\to \mathring m$ the diamond-shaped regions
disappear, and the resulting diagram consists of a string of
triangles. The Killing vector $\partial_t$ is then spacelike
everywhere, except on the degenerate horizons $\mathring r=r_b=r_c$.

On the level sets of $t$ a rather different analysis applies, this
is discussed in Section~\ref{sssXl2}.

\subsection{Generalised Kottler metrics} \label{ssgKm}

All the results discussed in Section~\ref{KSdS} remain valid if
$m\ne 0$ and if the metric $\mathring h$ in \eq{Schwm} is an
\emph{Einstein metric on an $(n-1)$--dimensional manifold $\Nnmo$
with scalar curvature equal to $(n-1)(n-2)$} \cite{Birmingham}. We
will refer to such metrics as \emph{generalised Kottler metrics}.
Note that $m=0$ requires $(\Nnmo,\zh)$ to be the unit round metric
if  one does not want $r=0$ to be a singularity at finite distance
along the level sets of $t$.

\subsection{Delaunay metrics}
 \label{sDDm}

The Delaunay metrics, in dimension $n\ge 3$, may be defined as the
(two parameter) family of metrics
\bel{Deq0}
 g= u^{4/(n-2)}(dy^2+\zh)
 \;,
\ee
where $\zh$ is the unit round metric on $\Sp^{n-1}$, which are
spherically symmetric and have constant scalar curvature $R(g) =
n(n-1)$. Thus the functions $u=u(y)>0$ must satisfy the ODE
\bel{Deq}
  u^{\prime\prime} -\frac{(n-2)^2}{4} u + \frac{n(n-2)}{4}
 u^{\frac{n+2}{n-2}} = 0.
\ee
The two parameters correspond respectively to a minimum value $\ve$
for $u$, with
\bel{epsr}
 0\leq \ve \leq \bar{\ve}=(\frac{n-2}{n})^{\frac{n-2}{4}}
\ee
($\ve$ is called the Delaunay parameter or neck size) and a
translation parameter along the cylinder.  A straightforward ODE
analysis (see~\cite{MPU1}) shows that all the positive solutions are
periodic.  The degenerate solution with $\ve = 0$ corresponds to the
round metric on a sphere from which two antipodal points have been
removed. The solution with $\ve = \bar{\ve}$ corresponds to the
rescaling of the cylindrical metric so that the scalar curvature has
the desired value.

Note that the Delaunay ODE was first studied by Fowler
\cite{Fowler1,Fowler2}, however the name used here and elsewhere in the
literature is  inspired from the analogy with
the Delaunay surfaces: the complete, periodic CMC surfaces of revolution in
$\R^3$~\cite{Del}. As is well known, the analogy between the
``conformally flat metrics of constant positive scalar curvature"
and ``complete embedded CMC surfaces of in $\R^3$" goes far beyond this
correspondence (see, e.g., \cite{MPo}).

Regarding the Delaunay metrics as singular solutions of the Yamabe
equation on $(\Sp^n, g_0)$ one has a number of uniqueness results.
Among these are the facts that no solution with a single singular
point exists, and that any solution with exactly two isolated
singular points must be conformally equivalent to a Delaunay metric.
These results can be proved by a generalization of the classical
Alexandrov reflection argument  (the method of moving planes),
see~\cite{GNN}.  The first general existence result for complete
conformally flat metrics of constant scalar curvature with
asymptotically Delaunay ends is due to Schoen~\cite{S88}.

Of immediate interest to us is the fact that  conformally flat
metrics, with constant positive scalar curvature, and with an
\emph{isolated singularity of the conformal factor} are necessarily
asymptotic to a  Delaunay  metric~\cite{KMPS};  {in fact, in
dimensions $n=3,4,5$ the conformal flatness condition is not
needed~\cite{Marques}}. Specifically, with respect to  {spherical}
coordinates about an isolated singularity of the conformal factor,
there is a half-Delaunay metric which $g$ converges to,
exponentially fast in $r$, along with all of its derivatives.  This
fact is used in~\cite{P93, MPU1,MPU2,MPa,Rat} where complete,
constant  scalar curvature metrics, conformal to the round metric on
$\Sp \setminus\{p_1,\ldots, p_k\}$ were studied and constructed.
(This is one instance of the more general ``singular Yamabe
problem".)

By the uniqueness of solutions to ODEs, or otherwise, we have:

\begin{Proposition} \label{funfact}
The time symmetric initial data sets for Kottler--Schwarzschild--de
Sitter space in spatial dimension $n$ with $\Lambda =
\frac{n(n-1)}{2}$, are precisely the Delaunay metrics with constant
positive scalar curvature $R=n(n-1)$.
\end{Proposition}

This correspondence continues to hold for any choice of positive
cosmological constant $\Lambda$ provided that one homothetically
rescales the Delaunay metrics so that $R=2\Lambda$.

Comparing \eq{Schwm} and \eq{Deq0} we find
\bel{ryc}
 r=u^{\frac 2 {n-2}}\;,\quad  r \frac {dy}{dr} = V^{-1/2} \;,
\ee
which allows us to determine $y$ as a function of $r$ on any
interval of $r$'s on which $V$ has no zeros.

\subsubsection{Extreme limit}
 \label{sssXl2}

Let $\mathring m$ and $\mathring r$ be as in Section~\ref{sssXl} and
suppose that $0<m<\mathring m$, denote by $r_*\in (r_b,r_c)$ the
value at which the maximum value $V_*$ of $V$ is attained, shifting
$y$ by a constant we can assume that  the corresponding value
$y_*=y(r_*)$ of the $y$ coordinate in \eq{Deq} is zero. We have
$r_*\to \mathring r$ and $V_*\to 0$ as $m\to \mathring m$, and it
clearly follows from \eq{ryc} that the correspondence
$y\leftrightarrow r$ breaks down in the limit. This singular
behavior with respect to the $r$ coordinate is of course resolved by
the coordinate $y$ of \eq{Deq0}. A somewhat more explicit way of
seeing this is to replace $r$ by a new coordinate $w$ through the
formula
$$
 r = r_*+ \sqrt{V_* }w
 $$
which scales up the interval $r\in (r_b,r_c)$ to  $w\in \left
((r_b-r_*)/\sqrt {V_*},(r_c-r_*)/\sqrt {V_*}\right)$. Equations
\eq{ryc} become
\bel{ryc2}
 r_*+\sqrt{V_* } w =u^{\frac 2 {n-2}}\;,\quad   \frac {dy}{dw} = \frac{\sqrt {V_*}}{(r_*+\sqrt{V_* } w)\sqrt V }
 \;,
\ee
which are regular in the limit $m\to \mathring m$. In the new
coordinates we have
$$
 \frac 1 V dr^2 + r^2 \mathring h =\frac {V_*} V dw^2 + (r_*+ \sqrt{V_* }w)^2 \mathring h
  \longrightarrow dw^2 +  r_* ^2 \mathring h\qquad\mbox{as} \qquad m\to\mathring m
  \;,
$$
with the   limit being uniform  over compact sets of the $w$
coordinate. This shows in which sense the space sections of the KSdS
metrics approach a cylindrical geometry in the extreme limit. It
should, however, be borne in mind that the space-time picture of
Section~\ref{sssXl} is rather different.

\subsection{Generalised Delaunay metrics} \label{ssgDm}

{Similarly to Section~\ref{ssgKm},}
the analysis presented at the beginning of Section~\ref{sDDm}
remains valid when the parameter $\epsilon$ is positive and if the
metric $\mathring h$ in \eq{Deq0} is an \emph{Einstein metric on an
$(n-1)$--dimensional manifold $\Nnmo$ with scalar curvature equal to
$(n-1)(n-2)$}. We   refer to the resulting metrics as
\emph{generalised Delaunay metrics}. {As before,} $\epsilon=0$ requires
$(\Nnmo,\zh\,)$ to be the unit round sphere if one wants to avoid a
singularity at the set $u(y)=0$.

\subsection{Complete metrics with constant positive scalar curvature
and asymptotically Delaunay ends}

Conformal gluing constructions for constant scalar curvature metrics
$\tilde{g}=u^{4/(n-2)}g$ with $R(\tilde{g})=R(g)=n(n-1)$ have given rise to a
wide variety of such metrics with asymptotically Delaunay ends. The
linearisation of this equation about a solution leads to the operator
$L_g=\Delta_g +n$.  The key {\em nondegeneracy} assumption of any
conformal gluing construction is that $L_g$ is surjective when acting on
appropriately defined function spaces. The Delaunay metrics themselves
are nondegenerate in this sense~\cite{MPU1}, moreover the solutions
constructed by Mazzeo-Pacard  on
$\Sp\setminus\{p_1,\ldots,p_k\}$~\cite{MPa} are non-degenerate. On the
other hand, the standard metric on the $n$-sphere, $(\Sp, g_0)$, is
degenerate due to the fact that the restrictions of the linear functions in
$\R^{n+1}$ span an $(n+1)$-dimensional co-kernel {of} $L_{g_0}$. In addition
to the original construction of Schoen~\cite{S88}, the constructions
of~\cite{MPU2} and~\cite{Rat} use  non-degenerate solutions as building
blocks to produce new non-degenerate solutions.

All of the constructions alluded to above are in the setting where
the metrics are locally conformally flat everywhere. This is clearly
not necessary.  A general conformal gluing theorem was established
by Byde~\cite{Byde}:

\begin{theorem}[Byde~\cite{Byde}] \label{byde}
Let $(M, g)$ be a compact Riemannian manifold, possibly with
boundary, of constant scalar curvature $n(n-1)$, which is
non-degenerate in the sense described above, and let $x_0\in
\rm{int}(M)$ be a point in a neighborhood of which $g$ is
conformally flat.  Then there is a constant $\rho_0$ and a one
parameter family of complete metrics $g_{\rho}$ on $M \setminus
\{x_0\}$ defined for $\rho\in (0, \rho_0)$, conformal to $g$, with
constant scalar curvature $n(n-1)$.   Moreover, each $g_{\rho}$ is
asymptotically Delaunay and $g_{\rho}\rightarrow g$  uniformly on
compact sets in $M \setminus \{x_0\}$ as $\rho\rightarrow 0$.
\end{theorem}

This result is exactly analogous to results for constant mean
curvature surfaces established in \cite{MPaPo1, MPaPo2}. Byde goes
further and shows how one can also glue asymptotically Delaunay ends
onto non-compact, non-degenerate solutions (though without the
uniform convergence to the original metric away from the gluing
locus).

Note  that, in light of Proposition~\ref{funfact}, all of these
results, and others, on the existence of CPSC metrics with
asymptotically Delaunay ends have an immediate reinterpretation,
after considering the maximal development of the initial  data set,
as statements regarding the existence of space-times with
asymptotically Kottler--Schwarzschild--de Sitter ends.

\section{Perturbation to exactly Delaunay, or generalised Delaunay ends}
 \ptc{a comma has been removed in the proofs in the title, please reinsert}

The gluing construction of
Corvino-Schoen~\cite{Corvino,CorvinoSchoen} (compare~\cite{ChDelay})
generalises  to the positive cosmological constant setting as
follows:

\begin{theorem}
 \label{TDel}
Let $\Nnmo$ be compact, let $(M, g)$ satisfy $R(g)=n(n-1)$, and
suppose that $M$ contains an end $E\approx [0,\infty)\times \Nnmo$
on which $g$ is asymptotic to a generalised Delaunay metric
$\zg=\zg_{\ve}$, with $0<\epsilon<(\frac{n-2}{n})^{\frac{n-2}{4}}$,
together with derivatives up to order four. Then for every
$\delta>0$ there is an $\ve^{\prime}$ satisfying $|\ve-\ve'|<\delta$
  and a metric
$g^{\prime}$ with $R(g^{\prime})=n(n-1)$, which differs from $g$
only far away on $E$, and which is a generalised Delaunay metric
with Delaunay parameter $\ve^{\prime}$ on the complement $E'$ of a
compact subset of $E$.
\end{theorem}

\begin{remark}
By taking the maximal, globally hyperbolic, space-time development of
the time-symmetric initial data set $(M, g^{\prime})$, we obtain a
solution of the vacuum Einstein equations with cosmological constant
$\Lambda = n(n-1)/2$   such that the metric on the domain of
dependence of the end $E^{\prime}$ is  isometric to a subset of the
Kottler--Schwarzschild--de Sitter space-time.
\end{remark}

\proof
Let $g$ asymptote to a generalised Delaunay metric $\zg_\varepsilon$
on $\R\times \Nnmo $. We can write $\zg_\varepsilon$ as
\bel{Schwm2}
 \zg_\epsilon= dx^2 + e^{2f(x)}\zh
 \;,
\ee
where $\zh$ is an Einstein metric on $\Nnmo $, normalised as
described above.

Consider a  connected component of the set on which  $V
>0$, 
 \ptc{the sign ``$>$" has vanished in the proofs, please reinsert}
 where $V$ is the function
appearing in \eq{Schwm} for the metric $\zg_\varepsilon$. It follows
from \eq{Schwm} that $\sqrt{V}$ is the normal component  of the
Killing vector $\partial_t$ on the level sets of $t$. From the
general results in~\cite{MR54:4541} it follows that any such
function, for a static space-time, solves  equation \eq{normKID}
below. It further follows from the analysis in~\cite{ChAPP} that
$\sqrt V$ can be smoothly continued to a real-analytic function on
$M$, which we call $\sqrtV$,  by changing signs across the zero
level sets of $V$. Furthermore, both $\sqrtV$ and $V=\pm \sqrtV^2$
are functions of $x$ only in the representation \eq{Schwm2} of
$\zg$.

Let $T=T(\varepsilon)$ be the period of $f$, and let
$\Omega_i=[iT+\sigma, iT+T+\sigma]\times \Nnmo $, where $\sigma$
will be chosen below.

Let $(\R \times \Nnmo , \zg_{\varepsilon'})$ be a generalised
Delaunay metric with parameter $\varepsilon'$ near $\varepsilon$,
$|\varepsilon-\varepsilon'|<\delta$.

Let $g_{\varepsilon'}$ be a metric on $\Omega_i$ obtained by
interpolating between $g$ and $\zg_{\varepsilon'}$ using any
$i$--independent cut-off function smoothly varying from zero to one.
The cut-off should be supported away from the end-points of the
interval $[i, i+T]$.

To {achieve} constant scalar curvature we will, first, correct the
metric $g_{\varepsilon'}$ to a new metric $\tg_{\varepsilon'}$ using
the operator $L$ of~\cite{ChDelay}, as restricted to time-symmetric
data, so that $Y\equiv 0$ there. The correction will be of the order
of the perturbation introduced, namely
$O(|\varepsilon-\varepsilon'|)$. This will, however, not quite solve
the problem because the operator $L$ at $g=\zg_\varepsilon$ has a
cokernel, which consists of functions solving the ``static
 \ptc{please change "normal KIDs" to ``static KIDs"}
KIDs equation":
\bel{normKID}
 D_iD_j N = N R_{ij} +\Delta_g N g_{ij}
 \;.
 \ee
We thus have to understand the space of solutions of \eq{normKID}:

\begin{Lemma}
\label{Londime} Let   $g=\zg$ be a generalised Delaunay metric as in
\eq{Schwm2}.
\begin{enumerate}
\item If $(\Nnmo,\mathring h)$ is the round sphere and if $m=0$, then the space of solutions of \eq{normKID} is
$(n+1)$--dimensional, spanned by $\sqrtV$ together with functions of
the form $e^{f} \alpha ^i$, where $f$ is as in \eq{Schwm2} and
$\alpha^i$ is the restriction of the Euclidean coordinate $x^i$ to
$\Snmo $ under the standard embedding $\Snmo \hookrightarrow R^n$.
\item Otherwise  all solutions of \eq{normKID} are proportional to
$\sqrtV$.
\end{enumerate}
\end{Lemma}

\proof Let $v^A$ be coordinates on the level sets of $x$; we have
$\Gamma^x_{AB}=-f' \zh_{AB}$, $\Gamma^x_{xj}=0$,
$\Gamma^A_{xB}=f'\delta^A_B$, and $\Gamma^A_{BC} = \zG^A_{BC}$,
where the $\zG^A_{BC}$'s are the Christoffel symbols of $\zh$. Since
$\sqrtV$ depends only upon $x$, and satisfies \eq{normKID}, we
immediately find
$$
 R_{xA}=0
$$
away from the zero-set of $V$; by continuity this holds everywhere
(this conclusion could also have been reached directly from the
warped product structure of $\zg$). But then \eq{normKID} gives
$$
 0=D_x D_A N = e^f \partial _x (e^{-f} \partial _A N)
 \;,
$$
hence \newcommand{\zP}{\mathring P}
\bel{angdep}
 N(x,v^A)=e^{f(x)} \zP (v^A)+ \zM(x)
 \;,
\ee
for some functions $\zP=\zP(v^A)$,  $\zM=\zM(x)$.

Set
$$
 \alpha:=N/\sqrtV
  \;,
$$
then $\alpha$ is smooth away from the zero-level sets of
$\sqrtV$. Since both $N$ and $\sqrtV $ satisfy \eq{normKID} one
finds that $\alpha$ is a solution of the equation
\bel{reduced0}
 \sqrtV  D_i D_j \alpha + D_i \alpha D_j \sqrtV  + D_j\alpha  D_i \sqrtV  =0
 \;.
\ee
{}From $D_A \sqrtV =0$ we obtain
\bel{reduced}
 D_A D_B \alpha = 0 \;.
\ee
Let $\lambda_{AB}=-\Gamma^x_{AB}= f' \zh_{AB}$ be the second
fundamental form of the level sets of $x$, \eq{reduced} can be
rewritten as
\bel{reduced2}
 \zD_{\! A}  \zD_{\! B}  \alpha+\alpha_x \lambda_{AB} \equiv
 \zD_{\! A}  \zD_{\! B}  \alpha+ \underbrace{\alpha_x f'}_{=:\varphi}\zh_{AB}=0
 \;,
\ee
where $\zD$ is the covariant derivative operator of the metric
$\zh$, and $\alpha_x=\partial_x \alpha$. Applying $\zD^B$ to
\eq{reduced2} and commuting derivatives one obtains (recall that the
Ricci tensor of $\zh$ equals $(n-2)\zh$)
$$
\zD_{\! A} \Big(\zD_{\! B} \zD^B \alpha + (n-2)\alpha -
\varphi\Big)=0\;.
$$
Contracting $A$ with $B$ in \eq{reduced2} we find $\varphi = \zD_{\!
B} \zD^B \alpha/(n-1)$, which allows us to conclude that there
exists a constant $C$ such that
\bel{difeqas}
 \zD_{\! B} \zD^B (\alpha +C)=- (n-1) (\alpha +C) \;.
\ee
Suppose, first, that $(\Nnmo,\zh)$ is the unit round sphere.
\Eq{difeqas}   shows that solutions of \eq{reduced2} are linear
combinations of the constant function $\alpha^0=1$ and of the
functions $\alpha^i=x^i|_{\Snmo }$, where $x^i$ is a canonical
coordinate in $\R^n$, with ${\Snmo }$ being embedded in $\R^n$ in
the obvious way. Hence, there exist functions $\lambda_\mu(x)$,
$\mu=0,\ldots,n$, such that
\bel{coc}
 \alpha(x,v^A) = \lambda_\mu(x) \alpha^\mu(v^A)
 \;.
\ee
But then \eq{angdep} implies $\lambda_i(x) = e^f\sqrtV ^{-1}
\zlambda_i$ for some constants $\zlambda_i$, without however
imposing any constraints on  $\lambda_0(x)$ which remains
undetermined so far.

On the other-hand, if $(\Nnmo,\zh)$ is \emph{not} the unit round
sphere, then by a theorem of Obata~\cite{Obata62} the function
$\alpha$ does not depend upon $v^A$, so that \eq{coc} again holds
with $\alpha^i\equiv 0$ and $\alpha^0=1$.

Inserting \eq{coc} into \eq{reduced0} with $ij=xx$ one finds
\bel{xxeq}
 \partial_x\Big\{\sqrtV^2 (x) \partial_x\Big[\lambda_0(x)+
 \sqrtV^{-1}(x)
e^{f(x)} 
 \zlambda_i \alpha^i(v^A)
  \Big]\Big\}=0
 \;.
\ee
In order to analyse this equation, it is useful to compare
\eq{Schwm} with \eq{Schwm2} to conclude that
\bel{dereq}
 \frac {dr}{dx}  = \pm\frac { 1}  {\sqrtV} = \pm\frac{ 1} { \sqrt{ 1 - \frac
 {2m}{ r^{n-2}} -\frac {r^2}{\ell^2}}}\;.
\ee
We then have $\sqrtV^2 \partial_x = \pm \sqrtV^3 \partial_r$ and
\beaa \pm \sqrtV^2  \partial_x\Big[\lambda_0 +
 \sqrtV^{-1}
e^{f } 
 \zlambda_i \alpha^i(v^A)
  \Big]
 &=&\sqrtV^3 \partial_r \Big[\lambda_0 + \frac r { \sqrt{1- \frac {r^2}{\ell^2}  - \frac
 {2m} { r^{n-2}}}} \zlambda_i \alpha^i(v^A)
  \Big]
  \\
  & = & \sqrtV^3 \partial_r \lambda_0  +1- \frac
 {mn} {r^{n-2}} \zlambda_i \alpha^i(v^A) \;.
\eeaa
So \eq{xxeq} will hold if and  only if this is a function which
depends at most upon $v^A$. Hence $\lambda_0$ is a constant and
$\zlambda_i=0$ unless $m=0$, in which case the $\zlambda_i$'s are
arbitrary, as desired.
 \qed

%
%
%
%
%
%
%
%
%

\bigskip

Returning to the proof of Theorem~\ref{TDel}, the metric $\tilde
g_{\varepsilon'}$ is obtained by solving the equation
$$
\Big(R(\tilde g_{\varepsilon'}) - n(n-1)\Big)\Big|_{\Omega_i}\in \
(\mathrm{Im }\,L)^\perp
$$
using the implicit function theorem,
compare~\cite[Theorem~5.9]{ChDelay}; this can be done on $\Omega_i$
for all $i$ large enough. As already mentioned, the perturbation
introduced is   $O(|\varepsilon-\varepsilon'|)$.  In view of
Lemma~\ref{Londime}, the obstruction to solving the problem is thus
the vanishing of
\bel{obstr}
 \int_{\Omega_i}  \sqrtV \Big(R(\tilde g_{\varepsilon'}) -R(\zg_{\varepsilon'})\Big)
 d\mu_{\zg}
 \;,
\ee
where $\sqrtV$ is the
 \ptc{please change "normal KID" to ``static KID"}
static KID associated with the generalised
Delaunay metric $\zg_{\varepsilon}$. We need the following identity,
from~\cite{ChHerzlich}:
\begin{eqnarray}
  \label{eq:3.2} &
  \sqrt{\det g} \;\sqrtV (R_g-R_b)  =  \partial_i \left(\ourU^i(\sqrtV )\right) + \sqrt{\det g}
\;(  \rho + Q)\;, &
\end{eqnarray}
where
\begin{eqnarray} \label{eq:3.3} & {}\ourU^i (\sqrtV ):=  2\sqrt{\det
g}\;\left(\sqrtV g^{i[k} g^{j]l} \zD_j g_{kl}
+D^{[i}\sqrtV  
g^{j]k} e_{jk}\right)
\;,&
\\
\label{eq:3.4} & \rho := (-\sqrtV \Ric(b)_{ij} +\zD_i \zD_j \sqrtV
-\Delta_b \sqrtV  b_{ij}) g^{ik} g^{j\ell}
  e_{k\ell}
\;,&
\\
\label{eq:3.5} & Q:= \sqrtV (g^{ij} - b^{ij} + g^{ik}g^{j\ell}
e_{k\ell})\Ric(b)_{ij} +Q'\;.
\end{eqnarray} Brackets over a symbol denote
anti-symmetrisation, with an appropriate numerical factor ($1/2$ in
the case of two indices). Here $ Q'$ denotes an expression which is
bilinear in
$$e\equiv e_{ij}dx^i dx^j
:= (g_{ij}-b_{ij})dx^i dx^j
 \;,
$$
and in $\zD_k e_{ij}$, where $\zD$ denotes  {now}  the covariant
derivative operator of the metric $b$, linear in $\sqrtV $, $d\sqrtV
$ and Hess$\sqrtV $, with coefficients which are constants in an ON
frame for $b$. The idea behind this calculation is to collect all
terms in $R_g$ that contain second derivatives of the metric in
$\partial_i \ourU^i$; in what remains one collects in $\rho$ the
terms which are linear in $e_{ij}$, while the remaining terms are
collected in $Q$; one should note that the first term at the
right-hand-side of \eq{eq:3.5} does indeed not contain any terms
linear in $e_{ij}$ when Taylor expanded at $g_{ij}=b_{ij}$. Note
that $\rho$ vanishes when $ b=\zg_{\varepsilon}$ by choice of
$\sqrtV$. So the integrand is quadratic in $e_{ij}$, up to terms
$O(|\varepsilon-\varepsilon'|) e_{ij}$, and up to the divergence
which produces a boundary term
$$
 \int_{\partial \Omega_i} \ourU^i dS_i
 \;.
$$

For our next lemma it is convenient to write two generalised
Delaunay metrics $g$ and $b$ as
\bel{gbeq}
g= \frac {dr^2} {N^2} + r^2 \zh\;, \quad
b= \frac {dr^2} {\zN^2} + r^2 \zh
\;.
\ee
We claim:

\begin{Lemma}
 \label{Lmassint}
Let $g$ and $b$ be two generalised Delaunay metrics with mass
parameters $m$ and $m_0$. Let $r$ be such that $\sqrtV (r)\ne 0$ and
$N(r)\ne 0$. If $\{r\}\times \Nnmo $ is positively oriented, then
\bel{massform}
 \int_{\{r\}\times \Nnmo } \ourU^i dS_i = 2\omega_{n-1} (n-1)
 \zN N^{-1}(m-m_0)
 \;.
\ee
where $\omega_{n-1}$ is the volume of $\Nnmo $.
\end{Lemma}

\proof  Let us denote by $\Gamma^i_{jk}$ the Christoffel symbols of
the metric $b$. We  have
$\Gamma^r_{rr}=-\partial_r \zN/\zN$, $\Gamma^r_{AB}=-r\zN^2 \zh_{AB}$,
$\Gamma^A_{BC}=\zGamma^A_{BC}$ (where, as before, the
$\zGamma^A_{BC}$'s are the Christoffel symbols of the metric $\zh$),
$\Gamma^A_{rB} = r^{-1} \delta^A_B$, while the remaining $\Gamma$'s
vanish.  It holds that $e= (N^{-2}-\zN^{-2})dr^2$, from which one
easily finds
\bel{massform2}
 \ourU^r  = 2 (n-1)  \zN N^{-1}(m-m_0) \sqrt{\det \zh}
 \;,
\ee
%
and the result follows by integration.
 \qed

We are ready to show  that one can choose $\varepsilon'$ ---
equivalently $m'$
--- so that the obstruction vanishes. So we consider the integral
\eq{obstr}. We wish to use \eq{eq:3.2} with $g=\tg_{\varepsilon'}$.
Note that the integration in \eq{obstr} is taken with respect to the
measure $d\mu_\zg$, while \eq{eq:3.2} involves $d\mu_g$.  The
difference between the two volume integrals comes thus with a
prefactor $O(|\varepsilon-\varepsilon'|)$, and produces an error
term which is $O\big((\varepsilon-\varepsilon')^2\big)$:
\bean
 \lefteqn
 { \int_{\Omega_i}  \sqrtV \Big(R(\tilde g_{\varepsilon'})
-R(\zg_{\varepsilon'})\Big)
 d\mu_{\zg}
 }
 &&
\\
\nonumber
 &&=
 \int_{\Omega_i}  \sqrtV \Big(R(\tilde g_{\varepsilon'})
-R(\zg_{\varepsilon'})\Big)
 d\mu_{g} + O\big((\varepsilon-\varepsilon')^2\big)
 \\
 &  &
 =
  \int_{\{i+T\}\times \Nnmo } \ourU^i dS_i  -  \int_{\{i\}\times \Nnmo } \ourU^i dS_i    + O\big((\varepsilon-\varepsilon')^2\big)
 \;.
\eeal{lastineq}
We now choose $\sigma$ in the definition of $\Omega_i$ so that  the
number
$$
 \lambda:=4\omega_{n-1} (n-1)
  \zN|_{ x=\sigma}
$$
does not vanish; note that $\lambda$ equals, up to
$O(|\varepsilon-\varepsilon'|)$, the number in front of $(m-m_0)$ in
\eq{massform}. Since $g$ approaches $\zg$ together with its first
derivatives as $i$ goes to infinity, the first integral in the last
line of \eq{lastineq} is  $ o(1) $, where  $o(1)$ tends to zero as
$i$ tends to infinity. By Lemma~\ref{Lmassint} the second integral
in the last line of \eq{lastineq} equals $\lambda(m'-\zm)=
O(|\varepsilon-\varepsilon'|)$, where $m'$ is the mass parameter of
$\g_{\varepsilon'}$ while $\zm$ is that of $\zg$. We infer that
$$
 { \int_{\Omega_i}  \sqrtV \Big(R(\tilde g_{\varepsilon'})
-R(\zg_{\varepsilon'})\Big)
 d\mu_{\zg} =\lambda(m'-\zm)+o(1) + O\big((\varepsilon-\varepsilon')^2\big)
} \;.
$$
Clearly this can be made positive or negative when $i$ is large
enough by choosing $\varepsilon'$ appropriately; by continuity there
exists an $\varepsilon'$ which makes the integral vanish, and the
result is proved. \qed

\section{Concluding remarks}
 \label{SCr}

Our work leads naturally to the following questions:

\begin{enumerate}
\item
In light of the results in \cite{Marques} one should expect that, at
least for $n=3, 4, 5$, a construction similar to Byde's~\cite{Byde}
could be carried out without any assumption of conformal flatness in
a neighborhood of the omitted point (which forms the end of the
resulting complete metric). Moreover, with or without the
conformally flat condition, it should be straightforward to iterate
Byde's construction to produce any number of  asymptotically
Delaunay ends. If one could add asymptotically Delaunay ends at any
chosen set of points in  any positive constant scalar curvature
manifold, then our analysis here could then be used to replace them
with exactly Delaunay ends. (Furthermore, all of this should be
doable via a \emph{local} deformation of the metric near points
\emph{without static KIDs}, using the techniques
of~\cite{ChBeignokids,CIP,ChDelay}.) Alternatively, can one
generically deform a constant  positive scalar curvature metric,
keeping the scalar curvature fixed, to a metric which is conformally
flat near a set of prescribed points? Proposition 4.1
of~\cite{SchoenCatini} could perhaps be used as an intermediate step
here.
%

\item In Byde's construction, or in a variation thereof as just
suggested, can one ensure that the range of masses of the resulting
Delaunay ends  covers an interval of the form $(0,\epsilon)$, for
some $\epsilon>0$?  This is a natural condition which has appeared
elsewhere as a necessary hypothesis (see e.g. \cite{Rat}).  Now, it
is clear that the masses of our exactly Delaunay ends  are
continuous functions of the initial mass. Given any two points with
the associated families of exactly Delaunay ends, one could then
always adjust the masses to be the same, ensuring that the ends can
be glued together.

\item
We  did not carry out  the gluing in situations when   the metric
$g$ approaches a cylindrical metric $dy^2 + \mathring h$ along the
asymptotic end; such metrics arise in black-hole space-times with
degenerate horizons. This deserves further attention.

\item
It would be of interest to extend the current gluings to general
relativistic initial data with non-vanishing extrinsic curvture.
 \end{enumerate}

\ptc{Hawking Gibbons ref \cite{GibbonsHawkingCEH} added; Riess ref \cite{Riess:2006fw}  updated and moved where it belongs from previous position 19; \cite{ChDelayAH} updated, \cite{Marques} updated}

\bibliographystyle{amsplain}
\bibliography{../references/bartnik,%
../references/myGR,%
../references/newbiblio,%
../references/newbib,%
../references/reffile,%
../references/bibl,%
../references/Energy,%
../references/hip_bib,%
../references/dp-BAMS,%
../references/MLproposal,%
../references/prop,%
../references/besse,%
../references/netbiblio
}%

\end{document}